\documentclass[10pt,letterpaper]{article}
\usepackage[top=0.85in,left=2.75in,footskip=0.75in]{geometry}

\usepackage{amsmath,amssymb}

\usepackage{changepage}

\usepackage{textcomp,marvosym}

\usepackage{cite}

\usepackage{nameref,hyperref}

\usepackage[right]{lineno}

\usepackage[nopatch=eqnum]{microtype}
\DisableLigatures[f]{encoding = *, family = * }

\usepackage[table]{xcolor}

\usepackage{array}

\usepackage{multirow}
\usepackage{booktabs}

\newcolumntype{+}{!{\vrule width 2pt}}

\newlength\savedwidth

\raggedright
\usepackage[aboveskip=1pt,labelfont=bf,labelsep=period,justification=raggedright,singlelinecheck=off]{caption}

\makeatletter
\renewcommand{\@biblabel}[1]{\quad#1.}
\makeatother

\usepackage{lastpage,fancyhdr,graphicx}

\usepackage{epstopdf}

\usepackage{siunitx}
\usepackage{textcomp}

\fancyheadoffset[L]{2.25in}
\fancyfootoffset[L]{2.25in}
\begin{document}
\vspace*{0.2in}

\begin{flushleft}
{\Large
\textbf{Characterization of DLBCL cell of origin-phenotypes based on tumor microenvironment features}
}
\newline
\\
Stefano Ugliano\textsuperscript{1,2 \Yinyang},
Martim Dias Gomes\textsuperscript{1,2 \Yinyang},
Noémie Moreau\textsuperscript{3},
Marta Pistone\textsuperscript{4},
Marcel Kirchner\textsuperscript{5},
Alexandra Just\textsuperscript{5},
Adrian Georg Simon\textsuperscript{4},
Christian Kukat\textsuperscript{5},
Reinhardt Buettner\textsuperscript{4},
Katarzyna Bozek\textsuperscript{1,2,6*}
% with the Lorem Ipsum Consortium\textsuperscript{\textpilcrow}
\\
\bigskip
\textbf{1} Institute for Biomedical Informatics, Faculty of Medicine and University Hospital Cologne, University of Cologne, Cologne, Germany
\\
\textbf{2} Center for Molecular Medicine Cologne (CMMC), Faculty of Medicine and University Hospital Cologne, University of Cologne, Cologne, Germany
\\
\textbf{3} Direction de la Recherche, de l'Innovation et des Relations Internationales (DRIRI), Centre Hospitalier Guillaume Régnier, Rennes, France.
\\
\textbf{4} Institute for General Pathology and Neuropathology, University Hospital Cologne, Cologne, Germany
\\
\textbf{5} FACS \& Imaging Core Facility, Max Planck Institute for Biology of Ageing, Cologne, Germany
\\
\textbf{6} Cologne Excellence Cluster on Cellular Stress Responses in Ageing-Associated Diseases (CECAD), University of Cologne, Cologne, Germany
\bigskip

% Insert additional author notes using the symbols described below. Insert symbol callouts after author names as necessary.
% 
% Remove or comment out the symbols in the byline and the author notes below if they aren't used.
%
% Primary Equal Contribution Note
\Yinyang \: These authors contributed equally to this work.

% % Additional Equal Contribution Note
% % Also use this double-dagger symbol for special authorship notes, such as senior authorship.
% \ddag These authors also contributed equally to this work.

% % Current address notes
% \textcurrency Current Address: Dept/Program/Center, Institution Name, City, State, Country % change symbol to "\textcurrency a" if more than one current address note
% % \textcurrency b Insert second current address 
% % \textcurrency c Insert third current address

% % Deceased author note
% \dag Deceased

% % Group/Consortium Author Note
% \textpilcrow Membership list can be found in the Acknowledgments section.

% Use the asterisk to denote corresponding authorship and provide email address in note below.
* k.bozek@uni-koeln.de

\end{flushleft}

% For PLOS Medicine research article authors, please structure your abstract
% with "Background", "Method and Findings" and "Conclusion" sections per
% journal requirements.

% For PLOS Neglected Tropical Diseases research article authors, please
% structure your abstract with "Background", "Methodology", "Findings", and
% "Conclusion" sections per journal requirements.
%
\section*{Abstract}
Diffuse large B-cell lymphoma (DLBCL) is an aggressive form of non-Hodgkin lymphoma with a high recurrence rate.
The molecular profiling of DLBCL tumors culminated in several immunohistochemistry algorithms for prognostic stratification.
Among those, the Hans classifier is widely used for classifying DLBCL  into  germinal center B-cell–like (GCB) and non-germinal center/activated B-cell–like (non-GCB/ABC) subtypes.
The Hans classifier primarily evaluates protein expression of tumor-associated markers, however the tumor microenvironment (TME) of DLBCL includes a myriad of immune and stromal cells, cytokines, and extracellular matrix components that contribute to tumor growth, immune evasion, and recurrence rate.
Although the Hans classifier provides a practical method for subtype identification, incorporation of TME information may improve risk stratification and further refine patient groups.
Here, we present an unbiased deep learning-based approach to extract meaningful features from TME of DLBCL tumors for the automated processing and analysis of multiplexed images of a DLBCL patient cohort. 
Our pipeline quantifies a range of features that describe tumor sample cell composition, morphology, and its spatial organization.
We point to alterations in the proportions of several cell populations between GCB and ABC tumors including increased immune cell proportions of the ABC and its preferential interaction with the M2-macrophages.
Our analysis offers an in-depth characterization of the DLBCL subtypes and is exemplary of how our pipeline can be used for detailed quantitative analysis of a tumor and its subtypes.

% Please keep the Author Summary between 150 and 200 words. Use first person.
% PLOS ONE, PLOS Biology, PLOS Global Public Health, PLOS Mental Health, and PLOS Water authors please skip this step. Author Summary is not valid for submissions to these journals.

% For PLOS Medicine authors, please structure your author summary with answers to the following questions:
% Why was this study done?
% What did the researchers do and find?
% What do these findings mean?
%
\break
\section*{Author summary}
Diffuse large B-cell lymphoma is commonly classified into GCB and non-GCB/ABC subtypes using the Hans classifier.
The Hans classifier relies on tumor protein expression markers but does not fully capture the complexity of the tumor microenvironment, which contains a multitude of immune and stroma cells that impact tumor outcomes.
The present work proposes an unbiased deep-learning framework to automatically analyze multiplexed DLBCL tissue images and extract biologically meaningful information.
Integrating TME-derived features with traditional subtype classification may improve prognostic stratification and enable more precise identification of high-risk patient groups.

\clearpage
\newgeometry{top=0.85in, left=1in, right=1in, footskip=0.75in}
% \linenumbers

\section*{Introduction}

Diffuse large B-cell lymphoma (DLBCL) is the most common type of non-Hodgkin lymphoma, accounting for around 30--35\% of cases worldwide.
DLBCL usually affects older adults, and it may arise in lymph nodes or in organs outside the lymphatic system---including the gastrointestinal tract, skin, and brain.
An early diagnosis and a prompt treatment are critical to ensure a better outcome for DLBCL patients.
As far as 40\% of all patients will suffer relapse in the first two years after initial therapy completion.
The prognosis for those patients is generally poor, therefore understanding why relapses occur and how these cases differ from non-relapsing ones is of utmost importance.

 DLBCL tumors are highly heterogeneous, differing in both morphology and clinical outcomes.
 Previous studies have used DNA microarrays and immunohistochemistry to classify DLBCL into molecular subtypes.
 The Hans classifier \cite{hans2004confirmation} is one of several immunohistochemistry-based classification algorithms that probe tumors for three clinically relevant markers: CD10, MUM1, and bcl-6.
 According to this classifier, DLBCL tumors can be divided into two main subgroups, the \emph{germinal center B-cell–like} (GCB) and the \emph{non-GCB activated B-cell–like} (ABC), based on their cell of origin.
 Frequently, non-GCB tumors are associated with a poorer prognosis, especially when treated with standard chemotherapy \cite{nowakowski2015abc} \cite{reber2013determination}. 
Although it represents an incomplete molecular characterization of the tumor, the Hans classifier remains a commonly used surrogate method for DLBCL cell-of-origin determination in routine pathology as it is cost-effective and feasible on formalin-fixed tissue.
It is also considered a useful proxy for more complex molecular biology;
however, its prognostic and therapeutic implications must be interpreted in the context of other clinical and molecular findings \cite{shimkus2023molecular}. 

The recent development of quantitative tools to describe cell morphology, microenvironment composition, and clinical variability enabled to gain new prognostic insights into tumor biology.
These quantitative approaches often rely on image data sourced from image scans of paraffin-embedded patients' biopsies stained with hematoxylin and eosin (H\&E), a routine procedure done at the pathology department of most clinics \cite{Onkologie2022DLBCL}.
While this kind of data is pivotal for diagnostics, disease-scoring, and prognostics, it only provides a coarse perspective of the tissue architecture, in which many molecular and morphological details are missing.
More recently, the advent of high-throughput microscopic techniques\cite{lin2018highly} enabled the creation of high-resolution multiplexed datasets containing information for a large number of antigen targets.
This kind of data captures more detailed information about the molecular and functional aspects of cells in a tissue. 

Here, we develop a flexible, modular deep learning-based approach for the automated processing and analysis of multiplexed images of a DLBCL patient cohort.
In our dataset of 106 patient samples, we perform a detailed characterization of the tumor microenvironment (TME) of the two most common DLBCL molecular subtypes: GCB and ABC.
Our analysis comprises the detection and quantification of TME cell populations, the extraction of meaningful relationships, and a large morphometry/statistical analysis of those tumors.

\section*{Methods}
\subsection*{Sample Preparation}
\subsubsection*{DLBCL tissue cohort assembly}
Human specimens from DLBCL were provided by the Department of Pathology and the Department of General, Visceral, and Transplant Surgery of the University Hospital Cologne in compliance with 20-1393 and following the privacy, data, and tissue use agreements according to the institutional review board at the Medical Faculty of the University of Cologne.
These human specimens consist of several tissue microarray blocks containing a total of 106 DLBCL patient samples with up to 3 tissue cores per patient and 69 healthy lymph nodes as control tissue. 
Tumor biopsy punches of 1.2 mm diameter and paired control tissue were assembled in-house with a tumor microarray (TMA) Grand Master (3D Histech, Hungary) robot, into 8 receiver formalin-fixed paraffin embedded (FFPE) blocks.
We used a a Leica Histocore Autocut (Leica biosystems, Germany) to cut individual slices of \SI{3}{\micro\meter} thickness from the paraffin blocks via tissue sectioning.
Individual samples used in this study were fully anonymized.
The anonymization procedure was approved by the review board at the Medical Faculty of the University of Cologne.

\bgroup
\begin{figure}[h!]
    \includegraphics[width=1\textwidth]{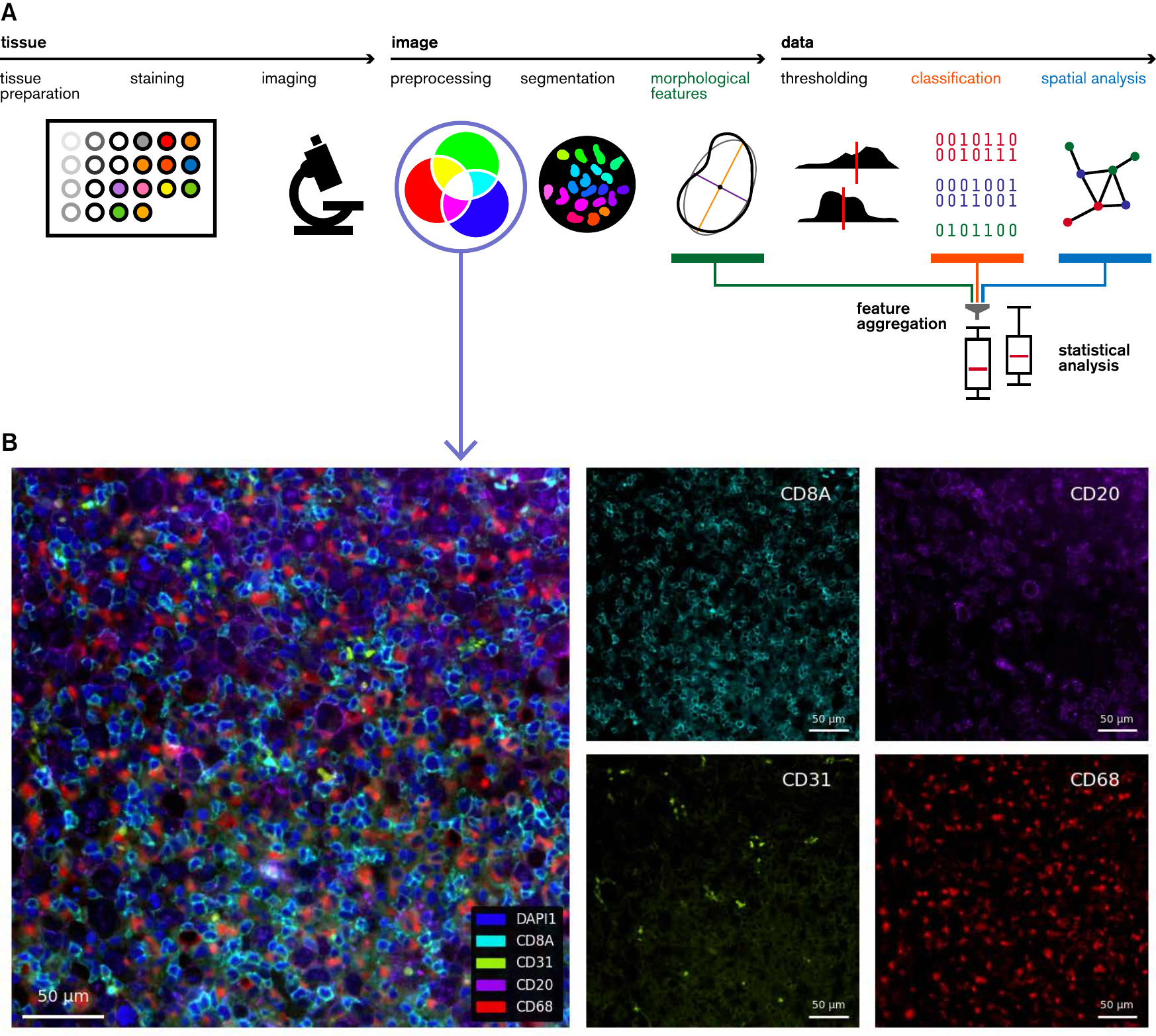}
    \caption{\textbf{A) Analysis pipeline.} Our pipeline consists of three stages: tissue preparation and imaging, image processing, and data extraction. Finally, we aggregate and analyze quantitative features of the ABC and GCB tumors in a statistical comparison.\\
    \textbf{B) Example tissue region and representative markers.} Left: Overlapping of DAPI and four markers in different colors. 
    Right: the four markers separately.
    This region was cropped from the center of a tissue sample.
    }
\label{fig:pipeline}
\end{figure}
\egroup

\subsubsection*{Tissue processing and staining}
Briefly, paraffin sections were deparaffinized, and antigens were retrieved in a tissue boiler (TissueMax, Germany) boiling in Citrate Buffer pH 6 for 20 min.
After PBS wash, blocking was performed for 1h with Interceptor Blocking Buffer (LI-COR Biosciences, Germany) at room temperature.
Fluorescence-conjugated antibodies (see Table~\ref{tab:panel}) diluted in Interceptor Blocking Buffer were applied overnight at 4°C in a humidified chamber.
DAPI staining was performed for 15 min before slide mounting with ibidi Mounting Medium (ibidi, Germany).
Before the first cycle of staining and after each round of imaging, slides were bleached with 4.5\% (wt/vol) $\mathrm{H_2O_2}$ and 20 mM NaOH in PBS.

\bgroup
\def\arraystretch{1.35}
\begin{table}[!ht]
\caption{\textbf{Staining antibodies.} Panel of antibodies used in our study.}
\centering
\small
\begin{tabular}{llllll}
\toprule
\textbf{Antibody}    & \textbf{Species}       & \textbf{Cell population}           & Fluorophore   & Company    & Catalog n. \\  \midrule %\hline \hline
CD8a        & human         & CD8+ T cells              & Alexa Fluor\texttrademark{} 488     & ThermoFischer & 53-0008-82   \\ %\hline
Lamin B     & human         & nuclear membrane          & Alexa Fluor\texttrademark{} 647     & abcam     & ab194108    \\ %\hline
CD31        & human         & endothelial cells/vessels & Alexa Fluor\texttrademark{} 555     & abcam     & ab279331    \\ %\hline
CD11b       & human         & dendritic cells           & Alexa Fluor\texttrademark{} 488     & ThermoFischer & 53-0196-80   \\ %\hline
CD3D        & human         & T cells                   & Alexa Fluor\texttrademark{} 555     & abcam     & ab208514    \\ %\hline
CD20        & human, monkey & B cells                   & eFluor 660    & ThermoFischer & 50-0202-82   \\ %\hline
CD163       & human         & macrophages               & Alexa Fluor\texttrademark{} 488     & abcam     & ab218293    \\ %\hline
CD68        & human         & macrophages               & Alexa Fluor\texttrademark{} 555     & abcam     & ab279323    \\ %\hline
CD204       & human         & macrophages               & Alexa Fluor\texttrademark{} 647     & abcam     & ab300100    \\ %\hline
CD4         & human         & CD4+ T cells              & Alexa Fluor\texttrademark{} 488     & R\&D Systems & FAB8165G    \\ %\hline
FOXP3       & human, monkey & T reg cells               & eFluor 570    & ThermoFischer & 41-4777-82   \\ %\hline
CD138       & human         & plasma cells, NK cells    & Alexa Fluor\texttrademark{} 647     & BioLegend   & 356523     \\ %\hline
Lamin A/C   & human         & nuclear membrane          & Alexa Fluor\texttrademark{} 488     & abcam     & ab185014    \\ %\hline
PD-L1       & human         & PD-L1                     & Alexa Fluor\texttrademark{} 555     & abcam     & ab274896    \\ %\hline
CD56        & human         & NK cells                  & eFluor 660    & ThermoFischer & 50-0565-80   \\ %\hline
\bottomrule
\end{tabular}
\label{tab:panel}
\end{table}
\egroup

\subsubsection*{Imaging}
We imaged stained slides in successive rounds of bleaching-staining-imaging.
The image acquisition was performed with a Zeiss Axioscan 7 fluorescence slide scanner equipped with a Zeiss Axiocam 712 Mono camera (Carl Zeiss Microscopy GmbH, Germany)
A custom protocol consisting of a 5x/0.25 NA low resolution fluorescence scan combined with an auto-threshold algorithm allowed for the automatic detection of individual tissue cores on the slide.
The resulting detection was then visually inspected and corrected when needed by manually drawing regions of interest around individual cores.
A 10x-magnification follow-up scan was used for autofocus setup at course (100 µm) and fine (10 µm) ranges.
% In order to ensure an optimal focal plan over a given region, 
We established several focus-points over the regions of interest, which were then used to find an optimal average focus plane. % over a given region.
Acquisition settings (light source intensity, exposure time and gain) were user-defined at the beginning of the acquisition for each cycle.
In total, five imaging cycles were performed with a 20x/0.8 NA Planapochromat objective using Colibri 7 light source (See Table~\ref{tab:channel_config} for details).

\bgroup
\def\arraystretch{1.35}

\begin{table}[h]
\small
\centering
\caption{\textbf{Microscope imaging channel configuration.}}
\label{tab:channel_config}
\begin{tabular}{rp{2.4cm}p{2.4cm}p{2.4cm}p{2.4cm}}
\toprule
 & \textbf{Channel 1} & \textbf{Channel 2} & \textbf{Channel 3} & \textbf{Channel 4} \\
\midrule
{Reflector} & 96 HE BFP    & 38 HE Green Fluorescence Prot & 43 HE DsRed  & 50 Cy 5     \\
                            % &               & Fluorescence Prot & & \\
Beam Splitter              & 420          & 495          & 570          & 660          \\
% Filter Ex. Wavelength (nm) & 370--410     & 450--490     & 538--562     & 625--655     \\
% Filter Em. Wavelength (nm) & 430--470     & 500--550     & 570--640     & 665--715     \\
% Contrast Method            & Fluorescence & Fluorescence & Fluorescence & Fluorescence \\
% Light Source               & LED-Module 385\,nm & LED-Module 475\,nm & LED-Module 567\,nm & LED-Module 630\,nm \\
Light Source Intensity     & 10\%      & 30\%      & 40\%      & 50\%      \\
Illumination $\lambda$ (nm) & 370--400   & 450--488     & 540--570     & 615--648     \\
% Channel Name               & DAPI         & AF488        & AF555        & AF647        \\
% Dye Name                   & DAPI         & AF488        & AF555        & AF647        \\
Excitation $\lambda$ (nm) & 353          & 493          & 553          & 653          \\
Emission $\lambda$ (nm)   & 465          & 517          & 568          & 668          \\
Effective NA               & 0.8          & 0.8          & 0.8          & 0.8          \\
Imaging Device              & 712m         & 712m         & 712m         & 712m         \\
%Camera Adapter             & 1x Camera Adapter & 1x Camera Adapter & 1x Camera Adapter & 1x Camera Adapter \\
Exposure Time (ms)         & 10           & 70           & 500          & 100          \\
Depth of Focus (\textmu m) & 1.45         & 1.62         & 1.78         & 2.09         \\
% Binning Mode               & 1,1          & 1,1          & 1,1          & 1,1          \\
% Scan Direction              & Unidirectional & -- & -- & -- \\
\bottomrule
\end{tabular}
\end{table}
\egroup

\bgroup
\def\arraystretch{1.2}
\begin{table}[!ht]
\caption{\textbf{Antibodies used in the following acquisition cycles.} The image channels from antibodies in parentheses were not further considered in the data analysis due to low quality of signal.}
\centering
\begin{tabular}{clcccc}
\toprule
\textbf{Cycle} &  & \textbf{Channel 1} & \textbf{Channel 2} & \textbf{Channel 3} & \textbf{Channel 4} \\ \midrule
1 &  & DAPI & CD8a & CD31 & (Lamin B) \\ 
2 &  & DAPI & CD11B & CD3D & CD20 \\ 
3 &  & DAPI & CD163 & CD68 & CD204 \\ 
4 &  & DAPI & CD4 & FOXP3 & (CD138) \\ 
5 &  & DAPI & -------- & PDL-1 & (CD56) \\ 
\bottomrule
\end{tabular}

\label{tab:cycles_antibodies}
\end{table}
\egroup

\subsection*{Image processing}

\subsubsection*{Image pre-processing and cell segmentation}
Raw data was stored in 65 .czi files, corresponding to 13 TMA slides imaged across 5 staining cycles.
Each TMA slide is composed of a variable number of cores, and each core was imaged in four different fluorescence channels, resulting in a four-channel image.
The image size of individual cores is approximately $16000\times16000$ pixels with a pixel size of \SI{172}{\nano\meter}, but image sizes differ due to the automatic stitching process of the individual tiles composing each image and to some variability in the shapes of the cores.

In order to improve the signal-to-noise ratio, we measured the background intensity from tissue-free areas (such as image corners) and subtracted it from the overall signal.
Finally, the corrected images were visually inspected for artifacts.

In all imaging cycles, the first channel (405nm) always corresponded to DAPI.
This reference staining was used to register the five different cycles, and to segment cell nuclei.
Registration of the cycles was obtained with a custom script based on pyStackReg, a Python wrapper of TurboReg/StackReg \cite{thevenaz1998pyramid}, while nuclear segmentation was performed using Cellpose \cite{pachitariu2022cellpose}.
We segmented DAPI images and compared the masks from five cycles to filter out those nuclei that do not appear across all cycles due to tissue damage or imaging problems.
We constructed the cytoplasm masks by expanding the nuclear masks by 25 pixels, or until the similarly expanded border of a neighboring next cell was met. Three channels were excluded from our analysis, CD138, PDL-1 and CD56 due to the poor quality of the antibody staining, high-background noise, unspecific-binding, and low signal-to-noise ratio.

\subsection*{Data Extraction}
We next extracted morphological and proteomic features at the single cell level from the nuclear/cellular segmentation masks.
The morphological features were computed directly from the masks, while the proteomic ones were obtained by quantifying protein levels within the respective cell and nuclei segmentation masks.

To quantify protein levels in a cell, we use pixel values within the respective protein expression site (nucleus, cytoplasm, both) and averaged them.
We next thresholded the single cell expression values with Otsu's algorithm \cite{otsu1975threshold} to obtain binary information---whether a cell is positive for a given marker or not.
The binarization is performed independently for each of the 11 markers, whereas the DAPI signal is not subject to thresholding.

\subsection*{Cell Classification}
\begin{figure}

\includegraphics[width=1\textwidth]{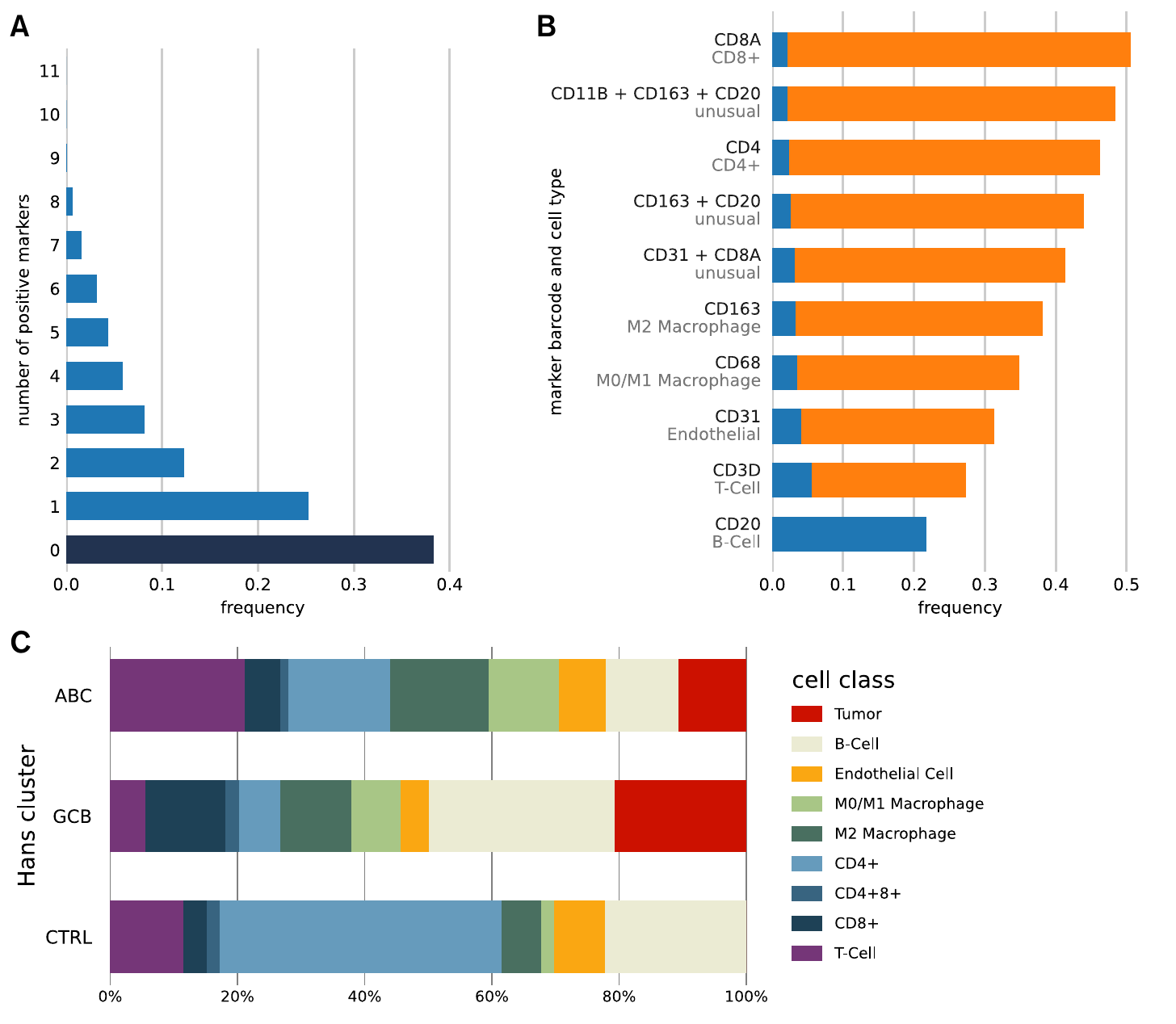}
\caption{
\textbf{A) Fraction of cells ($x$ axis) that are positive to a number of markers ($y$ axis).} Dark blue denotes cells that are negative to all markers.\\
\textbf{B) Top 10 most common marker combinations ($y$ axis, bottom to top) and the cell class they denote.}
Their fraction among all cells positive to any marker ($x$ axis) is shown in blue and the cumulative fraction in orange.\\
\textbf{C) Cell composition of Hans groups.} Proportion of cell classes in the tumor microenvironment of DLBCL groups according to Hans classifier, germinal center B-cell (GCB) and activated B-cell (ABC), CTRL (control).
}
\label{fig:comp2}
\end{figure}

After the binarization of the signal of the 11 marker channels, we label each cell according to the markers it is positive to.
Figure~\ref{fig:comp2}~A illustrates the fraction of cells that are positive with each number of markers.

Out of the $2^{11}$ possible marker combinations, only a few are observed  and the top 10 markers suffice to label 50\% of the cells (Fig.~\ref{fig:comp2}~B).
For example, a cell that is positive to CD31 is labeled as \emph{endothelial}, a cell that is positive to CD11b and CD163 is an \emph{M0/M1 macrophage}.
The labels and the respective combinations of markers are shown in a condensed form in Table~\ref{tab:classification_condensed}, and in an extended form in the Supporting Information.

\bgroup
\def\arraystretch{1.35}
\begin{table}[ht]
\caption{\textbf{Definition of cell types and their frequency.}
List of the marker combinations that we classified as specific cell types and their relative frequency across all segmented cells.
Presence of markers in brackets is optional, e.g.: a cell is labeled \emph{CD4+ T-helper cell} if it is positive to CD4, or CD4 and CD3D, or CD4 and PDL1, as \emph{endothelial} if it is positive to CD31, or CD31 and PDL1.
An extended version of this table is included in the Supporting Information.}
\centering
\begin{tabular}{rrl}
\toprule
\textbf{Cell type} & \textbf{Frequency} & \textbf{Marker combination} \\
\midrule
B cell NOS                      & 13.69\% & CD20 \\
CD4+ T-helper cell              &  2.60\% & CD4 + [CD3D or PDL1] \\
CD4+ activated T-helper cell    &  0.13\% & CD4 + CD11B + [CD3D] \\
CD8+ cytotoxic T cell           &  1.63\% & CD8A + [CD3D] \\
CD8+ activated cytotoxic T cell &  0.16\% & CD8A + CD11B \\
Double positive T cell          &  0.34\% & CD8A + [CD3D] \\
T cell NOS                      &  3.49\% & CD3D + [PDL1] \\
T cell NOS activated            &  0.15\% & CD3D + CD11B \\
T-regulatory cell               &  0.02\% & CD4 + FOXP3 + [PDL1] \\
Endothelial cell                &  2.55\% & CD31 + [PDL1] \\
M0/M1 macrophage                &  2.27\% & CD68 + [PDL1] \\
M2-macrophage                   &  3.47\% & \emph{(see Supporting Information: S1 Table)} \\
Monocyte NOS                    &  0.19\% & CD11B \\
PDL1                            &  0.27\% & PDL1 \\
\bottomrule
\end{tabular}
\label{tab:classification_condensed}
\end{table}
\egroup

A large proportion of cells (37.3\%) is not positive to any signal, 22.3\% are discarded as they express an unlikely combination of markers, and 9.5\% are left unclassified.
The remaining cells were categorized into cell types based on the protein markers (Table~\ref{tab:grouping}).
Different cell type groupings are defined, from finer to coarser.
In the finer grouping we distinguish 15 cell types, in a coarser one these 15 types are condensed into 10, and finally in the coarsest we group together different types of T-cells and macrophages, resulting in six cell types only.
The following analyses will focus on the middle classification with 10 cell types and the cells negative to all markers in our panel.

\bgroup
\def\arraystretch{1.35}
\begin{table}[ht]
\caption{\textbf{Cell types.} The cell types are defined in detailed categories and merged into groups of increasing coarseness.}
\centering
\begin{tabular}{lll}
\toprule
\textbf{Finer grouping} & \textbf{Coarser grouping} & \textbf{Coarsest grouping}  \\
\midrule
B cell NOS              & B cell                & B cell \\ \hline
CD4+ T-helper cell      & \multirow{2}{*}{CD4+} & \multirow{8}{*}{T cell} \\ \cline{1-1}
CD4+ activated T-helper cell &                  &  \\ \cline{1-2}
CD8+ cytotoxic T cell   & \multirow{2}{*}{CD8+} &  \\ \cline{1-1}
CD8+ activated cytotoxic T cell &               &  \\ \cline{1-2}
Double positive T cell  & CD4+8+                &  \\ \cline{1-2}
T cell NOS              & \multirow{3}{*}{T cell (other)} &  \\ \cline{1-1}
T cell NOS activated    &                       &  \\ \cline{1-1}
T-regulatory cell (T-reg) &                     &  \\ \hline
Endothelial cell        & Endothelial cell      &  Endothelial cell \\ \hline
M0/M1 macrophage        & M0/M1 macrophage      & \multirow{2}{*}{Macrophage} \\ \cline{1-2}
M2-macrophage           & M2-macrophage         &  \\ \hline
Tumor                   & Tumor                 &  Tumor \\ \hline
Monocyte NOS            & \multirow{3}{*}{other}& \multirow{3}{*}{other} \\ \cline{1-1}
PDL1                    &                       &  \\ \cline{1-1}
other                   &                       &  \\ 
\bottomrule
%Negative                & Negative              & Negative \\ \hline
%unusual                 & unusual               & unusual
\end{tabular}
\label{tab:grouping}
\end{table}
\egroup

\subsection*{Morphological Analysis}
The automatic segmentation allows to assign every pixel to the background or to one of the nuclei.
The shape of these nuclear masks carries information about the cell itself, its health and the replication stage it is currently in.
Moreover, cancer cell morphology has been also shown to be linked to the DLBCL response to treatment \cite{naji2025deep}.
We leveraged this morphological information and used it alongside the information extracted from the markers expression, namely the cell classification.
A list of the morphological features extracted is presented in Tab.~\ref{tab:feature_combinations}.

\subsection*{Spatial Organization of the DLBCLs}\label{sec:proximity}
We leverage the spatial information encoded in images in our analysis by constructing a proximity graph of each tissue sample.
In this graph, cells correspond to the nodes and edges exist between nodes if their two corresponding cells are in close proximity.

As we are missing a cell membrane marker in our imaging panel, we designed the following strategy to define cell proximity. 
For each segmented nuclear mask we know the coordinates of its centroid.
We average the major and minor axis of the moment ellipse of the nuclear mask to obtain the radius of the circular approximation of the nucleus.
Then for every pair of nuclei we compute the euclidean distance between the two centroids and subtract from this the sum of the two nuclear radii.
A tissue sample with $N$ segmented nuclei is represented by an $N\times N$ matrix of distances $D$ where element $D_{ij} = \sqrt{(x_i - x_j)^2 + (y_i - y_j)^2} - (r_i + r_j)$.
By choosing a threshold distance $d^*$ we define the adjacency matrix $A$ as \begin{equation*}A_{ij} = \begin{cases}
1 \text{ if } D_{ij} < d^*,\\0 \text{ otherwise}
\end{cases}
\end{equation*}
The choice of different values for $d^*$ produces different adjacency matrices, and therefore different graphs representing the tissue sample.
We chose the threshold at which two cells are considered adjacent at $d^*=0$ based on qualitative analysis of graphs at various cutoffs.

\subsection*{Feature Aggregation}
In order to compare different tissue samples, we aggregated the single cell data into vector representations that describe the entire tissue sample.
In the tissue description we consider morphological and spatial features of each cell type we defined based on the expression of respective protein markers.
Given a sample and one morphological feature $f\in F$, where $F$ is the set of all features, we obtain $\left|T\right|$ many distributions, one for each of the cell types in the set $T$.
We describe each of the statistical distributions with a series of values:
its mean, standard deviation, median, kurtosis, skewness, as well as its quantiles at chosen values \cite{sancere2025histo}.

This way for each sample we obtain tabular data in the shape of $N \times \left|F\right|$. 
We look at each of the cell types $t \in T$ independently and for each of those extract one distribution per morphological feature $f$.
Each of these distributions is then summarized in a vector of $\left|M\right|$ many metrics.
The final result is then a vector of length $\left|T\right| \cdot \left|F\right| \cdot \left|M\right|$ for each tissue sample.
More specifically, we include 10 cell types, 5 morphological features and 7 distribution metrics, resulting in a vector of length 350 per each sample.
These combinations are shown in Table~\ref{tab:feature_combinations}.

\begin{table}[h]
\caption{\textbf{Combination of components resulting in the feature vector.}
This feature vector describing each tissue sample is made of all possible combinations between all cell types with all morphological features and all statistical parameters used to describe their distribution.}
\label{tab:feature_combinations}
\vspace{4mm}
\begin{equation*}
\begin{array}{@{}l@{}}
  \text{B Cell}\\
  \text{CD4+}\\
  \text{CD4+8+}\\
  \text{CD8+}\\
  \text{Other T Cell}\\
  \text{Endothelial Cell}\\
  \text{M0/M1 macrophage}\\
  \text{M2-macrophage}\\
  \text{Tumor}\\
  \text{Other}\\
\end{array} \hspace{.4cm} \times \hspace{.4cm} \begin{array}{@{}l@{}}
  \text{Area}\\
  \text{Area Convex Hull}\\
  \text{Eccentricity}\\
  \text{Circular Approximation}\\
  \text{Extent}\\
\end{array} \hspace{.4cm} \times \hspace{.4cm} \begin{array}{@{}l@{}}
  \text{Mean}\\
  \text{Standard Deviation}\\
  \text{1st Quartile}\\
  \text{Median}\\
  \text{3rd Quartile}\\
  \text{Skewedness}\\
  \text{Kurtosis}
\end{array}
\end{equation*}
\end{table}

\subsubsection*{Additional Global Measures}
Other measures describing the cell types in the samples and the spatial proximity of these types are added to the 350 morphological features.
First, for each sample we calculate the proportions of the $\left|T\right|$ different cell types.
Then we consider the proximity graph corresponding to the sample, count the edges between all possible combinations of cell types and compute their proportions, resulting in additional $\frac{1}{2}\left|T\right| \cdot \left(\left|T\right|+1\right)$ features.

As a result, each sample is described by a vector of $\left(\left|T\right| \cdot \left|F\right| \cdot \left|M\right|\right)+ \left|T\right| + \frac{1}{2}\left|T\right| \cdot \left(\left|T\right|+1\right) = 350 + 10 + 110 = 470$ features.

\section*{Results}
\subsection*{Multiplexed immunofluorescence image data}
To capture the  highly heterogeneous microenvironment of DLBCL, we performed t-CyCiF multiplexed fluorescence imaging of 106 DLBCL patient samples. Our image data comprises 14 fluorescent markers in 559 tissue samples  (Fig.~\ref{fig:pipeline}~B).
In these data we identified 10 key cell populations (Table~\ref{tab:classification_condensed}) and quantified morphological, spatial and tissue organization features of the patient samples.

\subsection*{Altered microenvironment composition in DLBCL subtypes}
We first compared the cell composition of the two groups of samples according to Hans classifier and found several major differences between the composition of the TME of GCB and ABC tumors.
These differences include global proportions of endothelial, B-cells, macrophages, and T-cells (Fig.~\ref{fig:comp2}~C).
GCB samples are enriched in B-cells but show overall smaller proportions of macrophages and T-cells (Fig.~\ref{fig:comp2}~C).
These results suggest that the GCB TME is relatively less populated with immune cells compared to the ABC microenvironment.
Although macrophages are reduced as a bulk in GCB, the ratio of M0/M1 and M2-macrophages is increased in GCB compared to the ABC subtype.
Conversely, in the T-cell population, GCB tumors exhibit a smaller CD4+ population, expanded CD8+ population which contrasts with ABC tumors that show an almost inverted pattern with a larger CD4+ vs the CD8+ population (Fig.~\ref{fig:comp2}~C).
Additionally, a small but enriched CD4+ CD8+ T-cell population is present in the GCB and a large population of unspecific T-cells in the ABC subtype (Fig.~\ref{fig:comp2}~C).
Taken together, these results highlight differences in TME composition among GCB/ ABC tumors but also how differences in subpopulations are prototypical of a Hans subtype.

\subsection*{Distinct inter-cell population relationships in DLBCL subtypes}
\begin{figure}[h!]
\centering
  \includegraphics[width=1\textwidth]{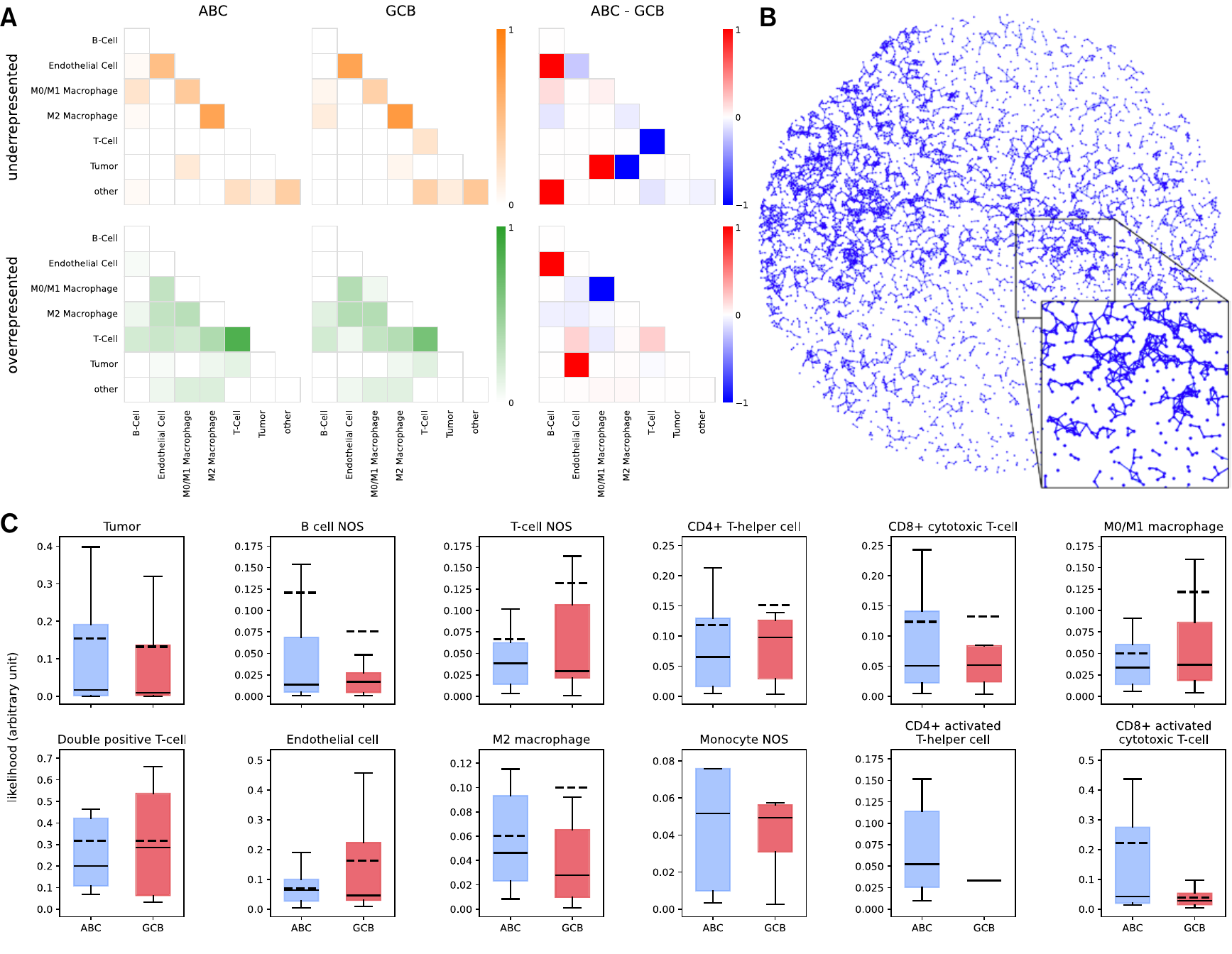}
  % \vspace{2mm}
 \caption{\textbf{A) Cell type interaction enrichment in the tumor microenvironment of DLBCL per Hans cluster group.}
 The color intensity in the heatmaps denotes the fraction of samples (per Hans class, as denoted by the column) in which cells of type $x$ and cells of type $y$ appear next to each other less (orange) or more often (green) in the respective samples compared to a randomized graph.
 That is, orange heatmaps show interactions of which pair of cell types are \emph{underrepresented in the tumor samples} compared to a randomized tissue graph, and the opposite for green heatmaps.
 Heatmaps in the right-most column depict how the cell type interactions differ between ABC and GCB samples.
 Red color indicates increased number of interactions of two cell types in ABC than GCB samples, and blue denotes the opposite.\\
 \textbf{B) Proximity graph.} A TMA core is represented as a graph with nodes corresponding to nuclei and edges joining nodes $A, B$ for which $d_{AB} \leq 0$ \\
 \textbf{C) Likelihood of cell neighborhood enrichment per Hans cluster group.} 
 Dashed lines indicate mean values and solid lines medians.
 }
\label{fig:comp3}
\end{figure}

To capture spatial organization of DLBCL tumors, we created proximity graph representations of the tumor tissues, where each cell is a node connected to its neighboring cells by an edge (Fig.~\ref{fig:comp3}).
These representations allow us to derive intra- and inter-cell populations relationships and neighborhood relationship between observations.
We first quantified global numbers of cell-cell interactions in both tumor subtypes.
We observed the highest number of interactions between cells of the same kind (Fig.~\ref{fig:comp3}).

To obtain a better insight into inter-cell interactions we devised a simulation to understand how likely are different cell types to be next each other in a random scenario.
Each tissue sample is represented by a graph, as explained in section ``Spatial Organization of the DLBCLs``:
nodes of the graphs are cells and edges and an edge connects two nodes if the corresponding cells are near in the tissue.
We labeled each node of this graph according to its cell type, and counted the number of edges across all pairs of cell types.
We then shuffled the node labels, while preserving the underlying graph structure, and counted the edges between pairs of cell types again.
By repeating this shuffling process we obtained a frequency distribution of the number of edges between each pair of cell types in a randomized graph. 
The observed values are compared to these distributions to determine which are over- or under-represented in the DLBCL tissue samples.
A number of interactions (such as Endothelial cells with B-cell, tumor cells with M0/M1 macrophages, ``other'' with B-cells, and Tumor with B-cells) were significantly enriched in ABC tumors vs GCB (Fig.~\ref{fig:comp3}).
Conversely, T-cell with T-cell interactions, Tumor with M2-macrophages and M0/M1 macrophages with M0/M1 macrophages were significantly depleted in ABC tumors vs GCB tumors (Fig.~\ref{fig:comp3}).

We then analyzed the different likelihoods of interaction of cells of any kind with the tumor cells.
Interestingly, we found that endothelial cells, M0/M1 macrophages and T-cells NOS were significantly more likely to interact with tumor cells in GCB tumors than ABC.
Conversely,  B-cells, CD8+ T-cells and M2-macrophages were more likely to interact with tumor cells in ABC subtype tumors than in GCB (Fig.~\ref{fig:comp3}).
These results suggest that, in addition to the different microenvironment composition in DLBCL subtypes, specific cell interactions are distinctive of both GCB and ABC tumors.

\subsection*{Statistical analysis of aggregated features}
\begin{figure}[!ht]
\centering
    \includegraphics[width=1\textwidth]{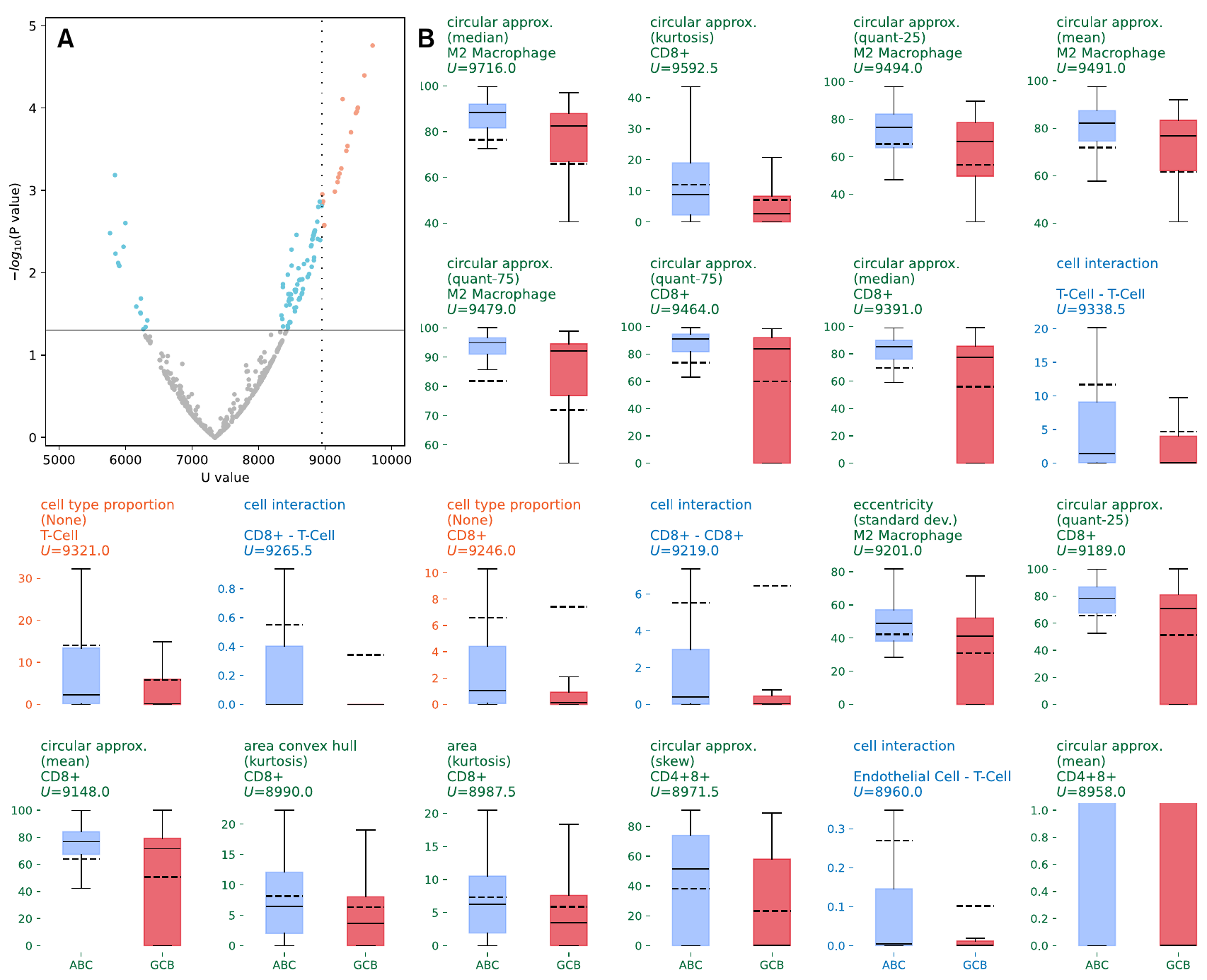}
\caption{\textbf{A) Feature significance.} The significance of all features (morphological, cell type frequency and proximity-likelihood) in distinguishing ABC vs GCB samples, according to Whitney $U$ test, visualized in a volcano plot.
The $x$ axis denotes the $U$ value while $y$ is a decreasing function of the p-value.
The horizontal line denotes the boundary of p-value $< 0.05$, so that points above it satisfy this criterion. 
The vertical dotted line shows the threshold value of Mann-Whitney $U$ delimiting the the top 20 metrics, marked in orange.\\
\textbf{B) Most significant variables distinguishing Hans subtypes.} Different colors in the titles reflect different classes of variables: morphological features (green), cell proportions (orange), and connection likelihood (blue).
Mean values are denoted by dashed lines, medians by solid lines. Arbitrary units were chosen for the $y$ axes}
\label{fig:top_boxplots}
\end{figure}

In order to build a systematic overview of the most significant morphological, cell composition, and cell interaction features that distinguish ABC from GCB tumors, we performed feature aggregation and obtained a compact, quantitative  representation of each tumor sample. 
We calculated the significance of feature value differences between the two tumor groups by means of a Mann-Whitney U test. Figure~\ref{fig:top_boxplots}~ illustrates $U$ against $-\log_{10}(\text{p-value})$ of the feature values.

We ranked the most significant features according to their $U$ value and looked closely at the top 20 (Fig.~\ref{fig:top_boxplots}~B). 
Most of the features are related to cell morphology.
Among those, the circularity and eccentricity of M2-macrophages and of CD8+ T-cells were important descriptors in distinguishing ABC from GCB tumors.
Moreover, T-cells proportions of both NOS and CD8+ T-cells were distinct among tumor subtypes.
Lastly, tumor cell interactions with T-cells, endothelial and homotypical interactions between T-cells were all distinctive of ABC samples vs GCB.
These results highlight specific differences in the TME of DLBCL tumors in respect of their cell-of-origin classification and suggest specific single-cell level differences in both macrophages and T-cells but also in the organization of local neighborhood in these two tumor molecular subtypes.

\section*{Discussion}

In the present study, we present a flexible, modular pipeline for the analysis of multiplexed imaging data.
We generated and processed a high-content dataset of DLBCL biopsy tissues stained for several immune cell markers important to the microenvironment of DLBCL.
Using our custom approach we segmented cell nuclei and corresponding bodies, identified specific cell populations, including their neighborhood and spatial relationships, and extracted quantitative features that describe the TME in an unbiased way. 
Our approach generates an array of multi-dimensional features that we used to characterize the TME of DLBCL including specific cell population composition, their morphology and spatial organization. Based on this quantification we compare tumors of different cell of origin subtype. 

Our analysis of the composition of the DLBCL TME revealed alterations in the proportions of several cell populations between GCB and ABC tumors.
The enrichment of B-cells in GCB could reflect a higher infiltration of non-tumor B-cells in the GCB vs ABC TME.
Information in literature is sparse regarding the role of non-tumoral B-cells on the microenvironment of DLBCL tumors \cite{berhan2025diffuse, bakhshi2020genetic}.
However, given the stark differences between ABC and GCB subtypes, we speculate that perhaps the ratio of tumor to non-tumor B-cells could be a good prognostic indicator. 
Importantly, we defined tumors by finding the intersecting population positive to CD20 and which had a nuclear area at least two times bigger than mean of the overall CD20 population.
While nuclear area is common identifier in clinical contexts \cite{Onkologie2022DLBCL} given the area of each TMA dot and the biopsy sampling approach, it is not a certain predictor of the tumor/non-tumor status of a CD20 positive cell. 
The numbers of non-tumor CD20 positive cells are potentially relevant however should be interpreted cautiously.

In bulk cell proportions GCB and ABC tumors show an immune-poor, immune-rich phenotype, respectively.
While deriving specific immune activity from multiplexed imaging is hard, these results are in line with previous reports of large subsets of cold-GCB cases and hot-ABC cases \cite{Tumuluru,Higashi, Kotlov2021}.
We also found an enrichment of macrophages and specially M2-macrophages in ABC tumors.
M2-macrophages have been linked to poorer prognosis in DLBCL and to be characteristic of ABC tumors \cite{chapuy2018molecular, Joldes2025, Pollari2025}.
Surprisingly, the ratio of M0/M1 and M2 was more imbalanced in GCB.
Previous reports have found an enrichment of M2-macrophages in GCB microenvironment but their role is unclear, suggesting that the role of M2-macrophages in the TME of DLBCL is highly context-dependent \cite{Joldes2025, Pollari2025,RanAn}.
The rather atypical enrichment of CD8+ T cells in GCB vs ABC could suggest that some of the tumors we analyzed had a strong PDL-1 phenotype as reported previously \cite{LacartePDL1}.

We reported two very interesting observations that were highly specific to GCB and ABC tumors respectively.
The first one concerns a reduced CD4+ CD8+ T-cell population in GCB.
In literature the CD4+ CD8+ double positive T-cells have been described as mature T-cell populations connected to aggressive disease especially in double‑hit DLBCL (MYC and BCL2) \cite{Li2022}.
In our cohort however, only two samples showed concomitant MYC translocation and BCL2 mutation, suggesting that this was not the main mechanism behind the low numbers of double positive T-cells. 
Importantly, double positive CD4+ CD8+ T-cell populations have also been described in healthy subjects, suggesting that their role in tumor progression is rather unclear and shall be further investigated \cite{ALAM2025103757,biomedicines11102702, PAREL2004215}.
Our second observation is that we found a intriguing large population of unspecified T-cell NOS in the ABC subtype tumors.
This population could potentially represent a group of exhausted T-cells (cytotoxic or else) that we failed to identify with our imaging protocol and detection methods.
This would be in line with a large group of exhausted T-cells there were found in other reports \cite{Huerga2025}. 

In addition to identifying the main cell populations in GCB and ABC tumors we quantified inter-cell relationships in the tumor environment.
GCB and ABC tumors showed different cell global interaction profiles in this analysis.
We found that endothelial cells, M0/M1 macrophages and T-cells NOS are more likely to interact with tumors cells in GCB tumors.
The interaction with endothelial cells is circumstantial, both tumors are known to induced angiogenesis and it could reflect biological differences among subjects.
Further, we only use CD31 as a specific marker for endothelial cells but it is known to be expressed to lesser extent in monocytes, dendritic cells, etc.
The interaction between M0/M1 and GCB tumor cells is in line with previous reports on the enrichment of M0/M1 macrophages in GCB tumors and their anti-tumorigenic potential in the DLBCL \cite{Joldes2025, Pollari2025} and contrasts with ABC tumors which showed a stronger interaction with M2-macrophages.
Lastly, we also found that the T-cell NOS population in GCB was more likely to interact with tumor cells, suggesting that potentially, in comparison to ABC tumors, the T-cells NOS we found have a more T-reg phenotype.
The differences in likelihood of CD8+ T-cells interaction with tumor additionally reflect the immune-rich nature of the ABC tumors.

Our pipeline culminates in the selection of the top 20 most statistically significant metrics that describe the cell and tissue level differences between ABC and GCB.
Among all the metrics we analyzed, the morphometrics related to M2-macrophages and  CD8+ T-cells were the most significant.
While the interpretation of individual measurements is hard, it is interesting that our results point to differences in both these cell populations.
Research in the role of M2-macrophages in DLBCL is extensive but context-dependent. It is known that M2-macrophage polarization which is highly dependent on morphometric features, is important in disease initialization and sustenance, but conclusive data is elusive \cite{Serna2023, FERRANT2026}.
As for the CD8+ T-cells majority of the studies focused on proportions of cells, infiltration into the tissue, and their exhaustion state.
We speculate that some of the morphometric differences we observed are a proxy for some of the cell state differences described previously \cite{Greenbaum2019, roider2024multimodal,ZHENG2026}. 

In summary, we established a unbiased deep learning-based approach to derive meaningful cell information from multiplexed imaging data of tumors samples.
We processed a DLBCL patient cohort with our pipeline and identified specific differences in the TME of ABC and GCB tumor subtypes.
Overall, our approach can be used to extract meaningful features specific to clinical subgroups and further contribute to patient risk stratification in DLBCL. 
Furthermore, given the modular and scalable nature of our pipeline, it can be used for the study of TME of different type of tumors.

\newpage
\section*{Supporting information}

% Include only the SI item label in the paragraph heading. Use the \nameref{label} command to cite SI items in the text.
\paragraph*{S1 table}
\label{app:classification}
\textbf{Cell Type Classification} The following is an extended table with all combinations of markers and the type of cell they denote.

\bgroup
\centering
\def\arraystretch{1.4}
\begin{table}[!h]
\caption{Extended table with all combinations of markers}
\label{tab:extended_markers}
\small
\begin{tabular}[t]{ll}
\toprule
\textbf{Classification} & \textbf{Markers}  \\ \midrule 
B cell NOS & CD20 \\
\hline
\multirow{3}{*}{CD4+ T helper cell} & CD4 \\
 & CD4 + CD3D \\
 & CD4 + PDL1 \\
\hline
%\multirow{2}{*}{CD4+ activated T helper cell} & CD4 + CD11B \\
% & CD4 + CD11B + CD3D \\
\multirow{2}{*}{CD4+ activated T helper cell} & CD4 + CD11B \\
 & CD4 + CD11B + CD3D \\
\hline
\multirow{2}{*}{CD8+ cytotoxic T cell} & CD8A \\
 & CD8A + CD3D\\
\hline
{CD8+ activated cytotoxic T cell} & {CD8A + CD11B} \\ \hline
\multirow{2}{*}{Double positive T cell} & CD4 + CD8A \\
 & CD4 + CD8A + CD3D \\
\hline
\multirow{2}{*}{T cell NOS} & CD3D \\
& CD3D + PDL1 \\
\hline
T cell NOS activated & CD11B + CD3D \\
\hline
\multirow{2}{*}{T regulatory cell} & CD4 + FOXP3 \\
 & CD4 + FOXP3 + PDL1 \\
\hline
\multirow{2}{*}{Endothelial cell} & CD31 \\
 & CD31 + PDL1 \\
\hline
\multirow{2}{*}{M0/M1 macrophage} & CD68 \\
 & CD68 + PDL1 \\
\hline
\multirow{12}{*}{M2-macrophage} 
& CD163 \\
& CD204 \\
& CD163 + CD204 \\
& CD163 + CD11B \\
& CD204 + CD11B \\
& CD163 + CD204 + CD11B \\
& CD163 + CD204 + CD11B + PDL1 \\
& CD163 + CD68 \\
& CD204 + CD68 \\
& CD204 + CD11B + CD68 \\
& CD204 + PDL1 \\
 & CD163 + CD204 + PDL1 \\
\hline
Monocyte NOS & CD11B \\
\hline
PDL1 & PDL1 \\
\bottomrule
\end{tabular}
\end{table}
\egroup

\newpage
% \bibliography{dlbcl.bib}

\end{document}